\documentclass[12pt]{iopart}

\usepackage{iopams}
\usepackage{bm}
\usepackage{mathrsfs}
\usepackage[utf8]{inputenc} 
\usepackage{enumerate}
\usepackage{appendix}
\usepackage{dcolumn}
\usepackage{multirow}
\usepackage{float}
\usepackage{color}

\usepackage{cancel}
\usepackage[utf8]{inputenc}
\usepackage{hyperref}

\makeatletter
\long\def\@makefntext#1{\parindent 1em\noindent 
 \makebox[1em][l]{\footnotesize\rm$\m@th{\arabic{footnote}}$}%
 \footnotesize\rm #1}
\def\@makefnmark{\hbox{${\arabic{footnote}}\m@th$}}
\def\@thefnmark{\arabic{footnote}}
\makeatother
\begin{document}
\title[Back-reaction in canonical analogue black holes]{Back-reaction in canonical analogue black holes}

\author{Stefano Liberati,$^{1,2,3}$ Giovanni Tricella,$^{1,2,3}$ and Andrea Trombettoni$^{4,1,2,3}$}

\address{$^1$ SISSA - International School for Advanced Studies, via Bonomea 265, \\
\qquad 34136 Trieste, Italy.}
\address{$^2$ INFN Sezione di Trieste, via Valerio 2, Trieste, Italy.}
\address{$^3$ IFPU - Institute for Fundamental Physics of the Universe, Via Beirut 2, \\
  \qquad 34014 Trieste, Italy.}
\address{$^4$ Department of Physics, University of Trieste, Strada Costiera 11, \\
  \qquad 34151 Trieste, Italy.}

\ead{liberati@sissa.it, gtricell@sissa.it, andreatr@sissa.it}
\vspace{10pt}

%
%\begin{indented}
%\item[]11 March 2018; 12 April 2018; \LaTeX-ed \today
%\end{indented}
%

%-----------------------------------------------------------------------------------------------------
\definecolor{purple}{rgb}{1,0,1}
\newcommand{\red}[1]{{\slshape\color{red} #1}}
\newcommand{\blue}[1]{{\slshape\color{blue} #1}}
\newcommand{\purple}[1]{{\slshape\color{purple} #1}}
%-----------------------------------------------------------------------------------------------------

\begin{abstract}
We study the back-reaction associated with Hawking evaporation of an acoustic 
canonical analogue black hole in a Bose--Einstein condensate.
We show that the emission of Hawking radiation induces a local back-reaction 
on the condensate, perturbing it in the near-horizon region, and a global
back-reaction in the density distribution of the atoms.
We discuss how these results produce useful insights into the process of
black hole evaporation and its compatibility with a unitary evolution.

\bigskip

%%%%%%%%%%%%%%%%%%%%%%%%%%%%%%%%%%
%%%%%%    Keywords      %%%%%%%%%%%%%%%%%%%%%
%%%%%%%%%%%%%%%%%%%%%%%%%%%%%%%%%%
%\noindent{\sc Keywords}: Analogue Gravity; Bose--Einstein Condensation.  
%%%%%%%%%%%%%%%%%%%%%%%%%%%%%%%%%%
%%%%%%%%%%%%%%%%%%%%%%%%%%%%%%%%%%
%%%%%%%%%%%%%%%%%%%%%%%%%%%%%%%%%%

\end{abstract}

%%%%%%%%%%%%%%%%%%%%%%%%%%%%%%%%%%
%%%%%%    PACS      %%%%%%%%%%%%%%%%%%%%%%%
%%%%%%%%%%%%%%%%%%%%%%%%%%%%%%%%%%
%\pacs{%insert
%} 
%%%%%%%%%%%%%%%%%%%%%%%%%%%%%%%%%%
%%%%%%%%%%%%%%%%%%%%%%%%%%%%%%%%%%
%%%%%%%%%%%%%%%%%%%%%%%%%%%%%%%%%%
\vspace{2pc}

\maketitle

\def\tr{{\mathrm{tr}}}
\def\cof{{\mathrm{cof}}}
\def\pdet{{\mathrm{pdet}}}
\def\d{{\mathrm{d}}}

% For two-column output uncomment the next line 
% and choose [10pt] rather than [12pt] in the \documentclass declaration
%\ioptwocol
%

%%-----------------------------------------------------------------------------------------------------
\hrule
\tableofcontents
\markboth{Back-reaction in canonical analogue black holes}
\bigskip
\bigskip
\hrule
\bigskip

%---------------------------------------------------------------------
\section{Introduction}
Hawking's prediction of black hole radiation in 1974~\cite{Hawking:1974rv,Hawking:1974sw} has been a milestone in our understanding of gravitation, by linking it to quantum field theory, thermodynamics and information theory, with far reaching consequences.

The presence of a horizon has two important consequences. First of all, it determines the causal separation of a region of spacetime, containing information that is no longer accessible by observers outside. Moreover, it induces the phenomenon of emission of black-body radiation which has the effect of extracting mass from the black hole leading to evaporation, and therefore to the apparent loss of the information contained inside that region.

The production of radiation can be predicted through the tools of quantum field theory in curved spacetime~\cite{Fredenhagen:1989kr,Jacobson:2003vx,Helfer:2003va,Harlow:2014yka}, but the description of the evaporation process requires a consistent formulation of semiclassical gravity, in which the degrees of freedom of geometry interact with the quantum degrees of freedom of matter fields. From a geometric point of view, this process entails the shrinking of the radius of the horizon and therefore the raise of the radiation temperature.

Furthermore, Hawking radiation introduces the so-called transplanckian problem~\cite{Fredenhagen:1989kr,Jacobson:2003vx,Helfer:2003va,Harlow:2014yka}, i.e.,~the~fact that low-energy Hawking quanta at spatial infinity and late times can be traced back to ultra-high-energy modes close to the horizon.
This appears to be a problem of Hawking's derivation as the latter relies on the validity of quantum field theory on curved spacetime which would be obviously untenable at Planck energies.
The theory of general relativity is not valid at such scales, and settling this issue would require a knowledge of the ultraviolet completion of the theory, so rising the question about the robustness of Hawking radiation with respect to the details of the latter.

Unfortunately, we presently do not have the ability of settling these issues experimentally in gravitational systems, as measuring Hawking radiation is basically impossible for astrophysical objects. However, Unruh's realization that hydrodynamical systems can be used for simulating black hole spacetimes~\cite{Unruh:1980cg,Visser:1998qn}, paved the way for devising more general tabletop experiments in which the propagation of some perturbations on a given background can simulate classical and/or quantum field theory on a curved spacetime. This realization marked the dawn of the so-called Analogue Gravity research field~\cite{Barcelo:2005fc}. 

In particular, in these analogue models the correspondence of their mathematical equations with those of black holes, opens the possibility to reproduce in a lab the analogue of Hawking radiation in the presence of a horizon, and so to start addressing long-standing issues such as the aforementioned ones. Indeed, in these systems the robustness of Hawking radiation with respect to a well understood ultraviolet completion of the effective spacetime and quantum field theory was soon argued for on theoretical grounds~\cite{Jacobson:1991gr,Unruh:1994je,Finazzi:2011jd}, further stimulating the attempt to achieve a direct observation. 

Bose--Einstein condensates (BEC) have been among the most successful analogue models due to their intrinsic quantum nature, simplicity and experimental realizability~\cite{Garay:1999sk,Visser:2001fe,Barcelo:2003wu,Barcelo:2005fc}. In these systems, Bogoliubov quasi-particles are quantum excitations that propagate following the acoustic metric induced by the wavefunction of the condensate, and the atomic structure provides a natural regularization of the theory at short distances. As such, they have provided an ideal setting for the study of analogue Hawking radiation and cosmological particle production~\cite{PhysRevA.76.033616,Balbinot:2007de,Carusotto:2008ep,Weinfurtner:2008if,Weinfurtner:2008ns,Sindoni:2009fc,Carusotto2010,Finazzi:2010nc,Zapata:2011ze,Anderson:2013ux,Fabbri:2020unn,Dudley:2020toe}. Even more remarkably, strong experimental evidence for Hawking radiation in BEC analogue systems was recently presented~\cite{Lahav:2009wx,Steinhauer:2015ava,Steinhauer:2015saa,deNova:2018rld,kolobov2019spontaneous}, largely confirming the theoretical expectations (at least in some suitable regime of the analogue black hole dynamics) and quite convincingly settling the issue of the robustness against possible high-energy modifications of the effective spacetime.
In particular, in BECs this can be deduced from the realization that the analogue Hawking spectrum retains its standard form, as~long as the typical wavelengths of the Hawking quanta are sufficiently larger than the healing length, which~plays the role of the quantum gravity scale in these effective spacetimes.

These results have of course stimulated new studies, also regarding the interplay of the microscopical (the analogue quantum gravitational) dynamics and the emergent phenomenology of spacetime and quantum fields. In particular, it was argued by the present authors that the other big issue related to Hawking radiation, the information loss problem---the apparent loss of unitary evolution implied by this phenomenon---can be understood using again the analogue gravity insight~\cite{Liberati:2019fse}. Indeed, in an isolated BEC system unitarity must be preserved and it is so once the evaporation process is considered at the microscopical level, tracking down the generation of correlations between the atoms underlying the emergent spacetime and the Hawking quanta.

Although in analogue gravity the back-reaction of the quantum fields on the acoustic geometry would not follow the Einstein equations, the observation of
evaporation in analogue black holes could provide information over the interplay of quantum fields and classical geometry within a semiclassical scheme in a broad class of quantum gravity scenarios. If one sees this analogue gravity system as a toy model for emergent gravity, it could then give a physical intuition of how Hawking radiation can be pictured as a feature emerging from an underlying full quantum theory.

Therefore, seeking this kind of theoretical insights, as well as the possibility to test this understanding of the back-reaction in experimental realizations, attracted an increasing interest on the nature of the back-reaction in analogue Hawking radiation~\cite{Schutzhold:2005ex,Fischer:2005iy,Liberati:2019fse,Goodhew:2019tax}. For example, in recent experiments with surface waves on a draining vortex water flow~\cite{Goodhew:2019tax}, it has been shown that the back-reaction is indeed observable, and it is possible to measure the exchange of energy and angular momentum between the background flow and the waves incident on the horizon. In Bose--Einstein condensates, a next step in this direction should be achievable as it should be possible to observe the back-reaction of the Hawking Bogoliubov quasi-particles on the acoustic geometry---i.e., on the substratum of condensate atoms---and interpret it through the lens of the underlying atomic theory.

With this aim in mind in this paper we focus our attention on the back-reaction of the Hawking radiation on a canonical analogue black hole within a BEC analogue system characterized by a $\lambda\phi^{4}$ interaction~\cite{Visser:1997ux,Dolan:2010zza,Vieira:2014rva}. The canonical analogue black hole is for our study a remarkable geometry because it is the spherically symmetric stationary solution of the Gross--Pitaevskii equation with homogeneous atom number density. The spherical symmetry allows reduction of the problem to a time-radius problem that presents a horizon at the radius where the velocity of the stationary ingoing superfluid equals the speed of sound.

In what follows we first describe the acoustic metric in this system and how the solutions of the Klein--Gordon equation behave at the horizon, and we calculate the Hawking radiation. These are the basis for showing how the back-reaction exerted by the Hawking radiation can be studied and how it affects the geometry of the black hole. We then propose to study a regime in which the Hawking radiation would lead to a sizeable evaporation of the analogue black hole. Finally, we show how this evaporation corresponds, in the evolution of the system, to the depletion of atoms from the condensate to the non-condensed part.
 
\section{Analogue Gravity in Bose--Einstein Condensates}\label{sec:AG-BEC}

We start by giving a brief overview of analogue gravity with Bose--Einstein condensates. This~fixes the notation and provides results used in the following Sections. 

Consider the local $\lambda\phi^{4}$ theory for a non-relativistic complex scalar field ($\hbar \equiv 1$) in 3+1 dimensions, with the usual bosonic commutation relations. The dynamics of the field is described by the Schr\"odinger field equation
\begin{eqnarray}
&&i\partial_{t}\phi = -\frac{\nabla^{2}}{2m}\phi+\lambda\phi^{\dagger}\phi\phi+V_{\mathrm{ext}}\phi \, ,\\
&&\left[\phi\left(x\right),\phi^{\dagger}\left(y\right)\right]=\delta^{3}\left(x,y\right)\, ,\\
&&\left[\phi\left(x\right),\phi\left(y\right)\right]=0\, .
\end{eqnarray}

Here and in the following, unless explicitly stated otherwise, we assume all the operators in a given correlation function to be evaluated at same time and same position.

In the presence of Bose--Einstein condensation we can describe the field by splitting it in mean-field and quantum fluctuations as
\begin{eqnarray}
&&\phi=\left\langle \phi\right\rangle +\delta\phi \, ,\vspace{3pt}\\
&&\left[\delta\phi\left(x\right),\delta\phi^{\dagger}\left(y\right)\right]=\delta^{3}\left(x,y\right)\, ,\vspace{3pt}\\
&&\left[\delta\phi\left(x\right),\delta\phi\left(y\right)\right]=0\, .
\end{eqnarray}

At leading order, the mean-field is described by the solution of the Gross--Pitaevskii equation~\cite{pitaevskii2003bose}
\begin{eqnarray}
i\partial_{t}\left\langle \phi_{0}\right\rangle &=&-\frac{\nabla^{2}}{2m}\left\langle \phi_{0}\right\rangle +\lambda\overline{\left\langle \phi_{0}\right\rangle }\left\langle \phi_{0}\right\rangle \left\langle \phi_{0}\right\rangle +V_{\mathrm{ext}}\left\langle \phi_{0}\right\rangle \, .
\end{eqnarray}

The linearized quantum fluctuation follows the Bogoliubov--de Gennes equations, which are coupled to the mean-field equation. When the mean-field is described by the Gross--Pitaevskii equation, the linearized quantum fluctuation can be solved separately after the solution of the Gross--Pitaevskii equation is found:
\begin{eqnarray}
i\partial_{t}\delta\phi&=&-\frac{\nabla^{2}}{2m}\delta\phi+2\lambda\overline{\left\langle \phi_{0}\right\rangle }\left\langle \phi_{0}\right\rangle \delta\phi+\lambda\left\langle \phi_{0}\right\rangle \left\langle \phi_{0}\right\rangle \delta\phi^{\dagger}+V_{\mathrm{ext}}\delta\phi \, .
\end{eqnarray}

Instead, if we include the back-reaction of the quantum fluctuation in the dynamics of the condensate wavefunction, the equations for the mean-field and the quantum fluctuation cannot be separated. The modified Gross--Pitaevskii equation includes the anomalous terms $n$ and $m$~\cite{Proukakis_2008} and the Bogoliubov--de Gennes equations are consequently modified, thus depending on this better approximation of the mean-field instead of the previous expression
\begin{eqnarray}
i\partial_{t}\left\langle \phi\right\rangle &=& -\frac{\nabla^{2}}{2m}\left\langle \phi\right\rangle +\lambda\overline{\left\langle \phi\right\rangle }\left\langle \phi\right\rangle \left\langle \phi\right\rangle \left(1+2n+m\right)+V_{\mathrm{ext}}\left\langle \phi\right\rangle \, , \label{eq:GP+BR}
\end{eqnarray}

\begin{eqnarray}
n &=& \frac{\left\langle \delta\phi^{\dagger}\delta\phi\right\rangle }{\overline{\left\langle \phi\right\rangle }\left\langle \phi\right\rangle }\, ,\\
m &=& m_{R}+im_{I}=\frac{\left\langle \delta\phi\delta\phi\right\rangle }{\left\langle \phi\right\rangle \left\langle \phi\right\rangle } \, .
\end{eqnarray}

These expressions will be used in Section~\ref{sec:BReac}.

With the usual Madelung representation introducing number and phase variables, the real and imaginary part of the Gross--Pitaevskii equation can be reorganized in the two quantum Euler equations for the condensate
\begin{eqnarray}
\left\langle \phi_{0}\right\rangle &=&\left\langle \rho_{0}\right\rangle ^{1/2}e^{i\left\langle \theta_{0}\right\rangle } \, , \\
\partial_{t}\left\langle \rho_{0}\right\rangle &=&-\frac{1}{m}\nabla\left(\left\langle \rho_{0}\right\rangle \nabla\left\langle \theta_{0}\right\rangle \right) \, , \label{eq:continuity}\\
\partial_{t}\left\langle \theta_{0}\right\rangle &=&\left\langle \rho_{0}\right\rangle ^{-1/2}\frac{\nabla^{2}}{2m}\left\langle \rho_{0}\right\rangle ^{1/2}-\frac{1}{2m}\left(\nabla\left\langle \theta_{0}\right\rangle \right)\left(\nabla\left\langle \theta_{0}\right\rangle \right)-\lambda\left\langle \rho_{0}\right\rangle -V_{\mathrm{ext}} \label{eq:externalpotential}\, .
\end{eqnarray}

We can also apply the Madelung representation to the quantum fluctuations

\begin{eqnarray}
&&\frac{\delta\phi}{\left\langle \phi_{0}\right\rangle }=	\left(\frac{\rho_{1}}{2\left\langle \rho_{0}\right\rangle }+i\theta_{1}\right) \, ,\\
&&\rho_{1} = \left\langle \rho_{0}\right\rangle \left(\frac{\delta\phi}{\left\langle \phi_{0}\right\rangle }+\frac{\delta\phi^{\dagger}}{\overline{\left\langle \phi_{0}\right\rangle }}\right) \, ,\\
&&\theta_{1}=	-\frac{i}{2}\left(\frac{\delta\phi}{\left\langle \phi_{0}\right\rangle }-\frac{\delta\phi^{\dagger}}{\overline{\left\langle \phi_{0}\right\rangle }}\right) \, ,\\
&&\left[\theta_{1}\left(x\right),\rho_{1}\left(y\right)\right]=-i\delta^{3}\left(x,y\right)\, ,\\
&&\left[\theta_{1}\left(x\right),\theta_{1}\left(y\right)\right]=0\, ,\\
&&\left[\rho_{1}\left(x\right),\rho_{1}\left(y\right)\right]=0\, ,
\end{eqnarray}

and rewrite the Bogoliubov--de Gennes equations obtaining two coupled dynamical equations for $\theta_{1}$ and $\rho_{1}$

\begin{eqnarray}
\left(\partial_{t}+\frac{\left(\nabla\left\langle \theta_{0}\right\rangle \right)}{m}\nabla\right)\frac{\rho_{1}}{\left\langle \rho_{0}\right\rangle }&=&-\frac{1}{m\left\langle \rho_{0}\right\rangle }\nabla\left(\left\langle \rho_{0}\right\rangle \nabla\theta_{1}\right) \, ,\\
\left(\partial_{t}+\frac{\left(\nabla\left\langle \theta_{0}\right\rangle \right)}{m}\nabla\right)\theta_{1}&=&\frac{1}{4m\left\langle \rho_{0}\right\rangle }\nabla\left(\left\langle \rho_{0}\right\rangle \nabla\frac{\rho_{1}}{\left\langle \rho_{0}\right\rangle }\right)-\lambda\rho_{1} \approx\nonumber\\
&\approx&-\lambda\rho_{1} \label{eq:negl_QP} \, ,
\end{eqnarray}
from which we see that these two real quantum fields are one the conjugate of the other. If we consider the quantum pressure to be negligible we obtain Equation~(\ref{eq:negl_QP}), which provides easily workable linearized expressions: when the spatial derivatives of $\rho_{1}$ are negligible, {$\rho_1$} is proportional to the flux derivative of the field $\theta_{1}$. 

The Bogoliubov equations can be combined into a second order partial differential equation for the real scalar field $\theta_{1}$, which appears to be a massless scalar field propagating in a curved spacetime. The resulting equation takes the form of a Klein--Gordon equation for an acoustic metric, analogue to that of a curved spacetime. One gets

\begin{eqnarray}
0&=&\partial_{\mu}\sqrt{g}g^{\mu\nu}\partial_{\nu}\theta_{1}=\nonumber\\
&=&	-\left(\partial_{t}+\nabla\frac{\left(\nabla\left\langle \theta_{0}\right\rangle \right)}{m}\right)\frac{m}{\lambda}\left(\partial_{t}+\frac{\left(\nabla\left\langle \theta_{0}\right\rangle \right)}{m}\nabla\right)\theta_{1}+\nabla\left\langle \rho_{0}\right\rangle \nabla\theta_{1} \, .
\end{eqnarray}

From this equation we can directly extract the elements of the inverse of the metric tensor. We do so rewriting the coupling $\lambda$ in terms of the speed of sound $c_{s}=\sqrt{\lambda\left\langle \rho_{0}\right\rangle /m}$ of the condensate:

\begin{eqnarray}
\sqrt{g} &=& \frac{\left\langle \rho_{0}\right\rangle ^{2}}{c_{s}} \, ,\\
g^{tt} &=& \frac{c_{s}}{\left\langle \rho_{0}\right\rangle }\left(-\frac{1}{c_{s}^{2}}\right) \, ,\\
g^{ti} &=& \frac{c_{s}}{\left\langle \rho_{0}\right\rangle }\left(-\frac{1}{c_{s}^{2}}\frac{\partial^{i}\left\langle \theta_{0}\right\rangle }{m}\right) \, ,\\
g^{ij}&=& \frac{c_{s}}{\left\langle \rho_{0}\right\rangle }\left(\delta^{ij}-\frac{1}{c_{s}^{2}}\frac{\partial^{i}\left\langle \theta_{0}\right\rangle }{m}\frac{\partial^{j}\left\langle \theta_{0}\right\rangle }{m}\right) \, ,
\end{eqnarray}
which correspond to the metric tensor of components
\begin{eqnarray}
g_{tt} &=& \frac{\left\langle \rho_{0}\right\rangle}{c_{s}} \left(-c_{s}^{2}+\delta^{ij}\frac{\partial_{i}\left\langle \theta_{0}\right\rangle }{m}\frac{\partial_{j}\left\langle \theta_{0}\right\rangle }{m}\right) \, , \\
g_{ti} &=& \frac{\left\langle \rho_{0}\right\rangle}{c_{s}} \left(-\frac{\partial_{i}\left\langle \theta_{0}\right\rangle }{m}\right) \, ,\\
g_{ij} &=& \frac{\left\langle \rho_{0}\right\rangle}{c_{s}} \delta_{ij} \, .
\end{eqnarray}

This acoustic metric is analogous to the metric of a 3+1 spacetime in Painlevé--Gullstrand coordinates
\begin{eqnarray}
\mathrm{d}s^{2}&=&-\left(1-\delta_{ij}V^{i}V^{j}\right)c^{2}\mathrm{d}t^{2}+2\delta_{ij}V^{i}c\mathrm{d}t\mathrm{d}x^{i}+\delta_{ij}\mathrm{d}x^{i}\mathrm{d}x^{j} \, ,
\end{eqnarray}
up to the conformal factor $\left\langle \rho_{0}\right\rangle /c_{s}$. The role of the velocity $V$ is played by the superfluid velocity
\begin{eqnarray}
V^{i}&=&-\delta^{ij}\frac{\partial_{j}\left\langle \theta_{0}\right\rangle}{m} \, .
\end{eqnarray}

In the case of spherical symmetry with $\partial_{r}\left\langle \theta_{0}\right\rangle <0$ we are in the case of ingoing Painlev\'e--Gullstrand coordinates, which allow study of horizon-penetrating trajectories and the behavior outside a black hole horizon. The solutions for the null radial geodesics are one regular ingoing solution and one outgoing solution that peels off the horizon (at the radius where $\left|V\right|=1$)  
\begin{eqnarray}
0&=&-\left(1-V^{2}\right)u^{t2}+2\left|V\right|u^{r}u^{t}+u^{r2}\, ,\\
&\Downarrow& \nonumber\\
u^{t}&=&-\frac{u^{r}}{\left|V\right|\mp1}\, .
\end{eqnarray}

In the following we will make use of the useful notation
\begin{eqnarray}
h\left(r\right)&=&1-V^{2}\left(r\right) \, ,
\end{eqnarray}
since it is most convenient to describe the behavior at the horizon.

%---------------------------------------------------------------------
\section{Canonical Analogue Black Holes}\label{sec:CABH}
%---------------------------------------------------------------------
Among the acoustic metrics simulated with Bose--Einstein condensates, it is convenient for our investigation to consider the canonical analogue black hole.

A canonical analogue black hole could be obtained as the stationary solution of the Gross--Pitaevskii equation with spherical symmetry and with homogeneous atom number density. Due~to the spherical symmetry, it presents a horizon at the radius where the velocity of the ingoing flow is equal to the speed of sound. The stationarity of the background metric, of the mean-field, means that the deviations from this configuration should be understood as the effect of the phenomenology previously neglected: the subleading effect of the quasi-particles and their back-reaction. The~homogeneity of the atom density $\left\langle\rho_{0}\right\rangle$ means that the conformal factor $\left\langle \rho_{0}\right\rangle /c_{s}$ is a number, and can trivially be transformed away.

Taking a constant and homogeneous number density $\left\langle\rho_{0}\right\rangle$, and assuming $\left\langle\theta_{0}\right\rangle$ to be a function of the radius only, from Equations~(\ref{eq:continuity}) and (\ref{eq:externalpotential}) we obtain the stationary spherically symmetric solution of the Gross--Pitaevskii equation. The superfluid velocity and the external potential setting the corresponding velocity profile are respectively

\begin{eqnarray}
-\frac{\partial_{r}\left\langle \theta_{0}\right\rangle }{m}=c_{s}\frac{r_{H}^{2}}{r^{2}}=V_{0} \label{eq:definition_V0} \, , \\
V_{\mathrm{ext}}=-m c_{s}^{2}\left(1+\frac{1}{2}\frac{r_{H}^{4}}{r^{4}}\right)\, ,
\end{eqnarray}
with an ingoing flux of atoms crossing every closed surface containing the origin at a rate of $4\pi c_{s}r_{H}^{2}\left\langle \rho_{0}\right\rangle $ atoms per unit time, which must be pumped in the system and move towards the singularity in the origin. The radius of the horizon $r_{H}$ is set by the external potential $V_{\mathrm{ext}}$.

Omitting the conformal factor, the analogue metric implied by Equation~(\ref{eq:definition_V0}) is therefore
\begin{eqnarray}
\mathrm{d}s^{2}=&-\left(1-\frac{r_{H}^{4}}{r^{4}}\right)c_{s}^{2}\mathrm{d}t^{2}+2\frac{r_{H}^{2}}{r^{2}}c_{s}\mathrm{d}t\mathrm{d}r+\mathrm{d}r^{2}+r^{2}\mathrm{d}\theta^{2}+r^{2}\sin^{2}\theta\mathrm{d}\phi^{2} \, .
\end{eqnarray}

The singularity in the origin poses a problem that would
be unavoidable in devising an experiment: one should either remove the
atoms reaching the origin and put them back in the system at large radius~\cite{Garay:2000jj}, or should consider only the region in which the
hydrodynamical limit holds with a good approximation, at a distance
from the origin and for a period of time such that it is not affected by atoms which flowed towards the origin.~(To implement the latter, one could devise a $3D$ hard wall trapping potential, possibly spherical, plus a variation in the central region of the scattering length and suitable external potential there added via a superimposed laser beam.)

These are however practical concerns that we shall assume dealt with in our theoretical study. In addition to them, we observe that the possible
removal of atoms from the system would affect the atomic correlation
functions and introduce statistical mixture, but the dynamics of the
quasi-particles in the near-horizon region, on which we focus on the
rest of the paper, would not be sensibly affected.

\subsection{Klein--Gordon Equation and Field Modes}\label{sec:KGeq}
%---------------------------------------------------------------------
The massless scalar field living in our analogue spacetime can be described in terms of its modes, found by solving the Klein--Gordon equation. For a stationary spherically symmetric spacetime in Painlev\'e--Gullstrand coordinates, the equation to solve is 
\begin{eqnarray}
\fl \quad
0&=&\frac{1}{\sqrt{g}}\partial_{\mu}\sqrt{g}g^{\mu\nu}\partial_{\nu}f\left(t,r,\Omega\right)=\nonumber\\
\fl \quad
&=&\left(-\partial_{c_{s}t}^{2}+\sqrt{1-h}\partial_{r}\partial_{c_{s}t}+\frac{1}{r^{2}}\partial_{r}r^{2}\sqrt{1-h}\partial_{c_{s}t}+\frac{1}{r^{2}}\partial_{r}r^{2}h\partial_{r}-\frac{L^{2}}{r^{2}}\right)f\left(t,r,\Omega\right) \, , \label{eq:KG}
\end{eqnarray}
where $\Omega$ is a compact notation for the solid angle and $L^{2}$ is
the usual angular momentum operator
\begin{eqnarray}
L^{2}=&-\frac{1}{\sin\theta}\partial_{\theta}\sin\theta\partial_{\theta}-\frac{1}{\sin^{2}\theta}\partial_{\phi}^{2} \, .
\end{eqnarray}

We can find a complete set of solutions looking for those which preserve
the symmetries of the system~\cite{Dolan:2010zza,Vieira:2014rva}, i.e.,~the eigenfunctions of both the time derivative and the angular
momentum operator
\begin{eqnarray}
f_{\omega lm}\left(t,r,\Omega\right)&=&e^{-i\omega t}Y_{lm}\left(\Omega\right)f_{\omega l}\left(r\right)\, ,
\end{eqnarray}
\begin{eqnarray}
\partial_{t}f_{\omega lm}\left(t,r,\Omega\right)&=&-i\omega f_{\omega lm}\left(t,r,\Omega\right)\, ,\\
L^{2}f_{\omega lm}\left(t,r,\Omega\right)&=&l\left(l+1\right)f_{\omega lm}\left(t,r,\Omega\right)\, ,\\
\partial_{\phi}f_{\omega lm}\left(t,r,\Omega\right)&=&imf_{\omega lm}\left(t,r,\Omega\right)\, ,
\end{eqnarray}
where the functions $Y_{lm}\left(\Omega \right)$ above are the usual spherical harmonics.

So Equation~(\ref{eq:KG}) becomes  
\begin{eqnarray}
\fl \quad
0&=&\left(\frac{\omega^{2}}{c_{s}^{2}}-i\frac{\omega}{c_{s}}\sqrt{1-h}\partial_{r}-i\frac{\omega}{c_{s}}\frac{1}{r^{2}}\partial_{r}r^{2}\sqrt{1-h}+\frac{1}{r^{2}}\partial_{r}r^{2}h\partial_{r}-\frac{l\left(l+1\right)}{r^{2}}\right)f_{\omega l}\left(r\right)=\nonumber\\
\fl \quad
&=&\frac{e^{i\frac{\omega}{c_{s}}\int^{r}\frac{\sqrt{1-h}}{h}\mathrm{d}r}}{rh}\left(h\partial_{r}h\partial_{r}+\frac{\omega^{2}}{c_{s}^{2}}-\frac{h\left(\partial_{r}h\right)}{r}-\frac{hl\left(l+1\right)}{r^{2}}\right)re^{-i\frac{\omega}{c_{s}}\int^{r}\frac{\sqrt{1-h}}{h}\mathrm{d}r}f_{\omega l}\left(r\right) \label{eq:KG_reorg} \, .
\end{eqnarray}

We have written Equation~(\ref{eq:KG_reorg}) reorganizing the various contributions to have a clearer view of the problem. The operator $h \partial_{r}$ can be understood as the directional derivative with respect to the tortoise radial coordinate. The term $h \partial_{r}h/r$ is a potential which induces mode mixing. This second order differential equation will have two linearly independent solutions for each set of eigenvalues, an~outgoing mode and an ingoing mode. Splitting the solutions in norm and phase we get two coupled real differential equations

\begin{eqnarray}
\fl\quad
\quad f_{\omega l}\left(r\right)&=&\sqrt{n_{\omega l}\left(r\right)}e^{i\phi_{\omega l}\left(r\right)}e^{i\frac{\omega}{c_{s}}\int^{r}\frac{\sqrt{1-h}}{h}\mathrm{d}r} \, ,\\
\fl\quad
\qquad\quad 0&=&\frac{1}{r}\left(h\partial_{r}h\partial_{r}+\frac{\omega^{2}}{c_{s}^{2}}-\frac{h\left(\partial_{r}h\right)}{r}-\frac{l\left(l+1\right)h}{r^{2}}\right)r\sqrt{n_{\omega l}\left( r \right)}e^{i\phi_{\omega l}\left(r\right)}\, ,\\
\fl\quad
&\Downarrow&\nonumber\\
\fl\quad
\partial_{r}\phi_{\omega l}\left(r\right)&=&\frac{C}{r^{2}hn_{\omega l}\left(r\right)} \, ,\\
\fl\quad
\qquad\quad 0&=&\frac{1}{r}\left(h\partial_{r}h\partial_{r}+\frac{\omega^{2}}{c_{s}^{2}}-\frac{h\left(\partial_{r}h\right)}{r}-\frac{l\left(l+1\right)h}{r^{2}}\right)r\sqrt{n_{\omega l}}-\frac{C^{2}}{r^{4}}\frac{1}{\sqrt{n_{\omega l}^{3}\left(r\right)}} \, ,\label{eq:KG_norm}
\end{eqnarray}

Different integrating constants $C$ set the sign of the derivative of the phase and give the different~solutions.

\subsection{Near-Horizon Behavior}

We are interested in the solutions in proximity of the horizon, for a small distance $\delta r = r-r_{H}$, and such that $n_{\omega l}\left( r\right)$ is regular. The horizon-crossing solutions will have a regular behavior in phase, and the horizon-tangent solutions will be the remaining linearly independent solutions. In the presence of a horizon, $h$ vanishes at $r_{H}$ and so do all the terms depending on it in the Klein--Gordon equation. We~can therefore set the phase-defining constant $C$ of the solutions with regular norm by requiring it to satisfy the Klein--Gordon equation at the horizon
\begin{eqnarray}
&&h\left(r_{H}+\delta r\right)=h^{\left(1\right)}\left(r_{H}\right)\delta r+\mathcal{O}\left(\delta r^{2}\right)%=4\frac{\delta r}{r_{H}}+\mathcal{O}\left(\delta r^{2}\right)
\, , \\
&&n_{\omega l}\left(r_{H}+\delta r\right)=n_{\omega l}\left(r_{H}\right)+\mathcal{O}\left(\delta r\right)\, ,\\
&&C=\pm\frac{\omega}{c_{s}}r_{H}^{2}n_{\omega l}\left(r_{H}\right) \, .
\end{eqnarray}

Considering the phase of the function $f_{\omega l} \left( r \right)$ we get, in the specific case of the canonical analogue black hole,
\begin{eqnarray}
\partial_{r}\arg\left(f_{\omega l}\left(r\right)\right)&=&\frac{C}{r^{2}hn_{\omega l}\left(r\right)}+\frac{\omega}{c_{s}}\frac{\sqrt{1-h}}{h}=\nonumber\\
&=&\pm\frac{\omega}{c_{s}h}\left(\frac{r_{H}^{2}n_{\omega l}\left(r_{H}\right)}{r^{2}n_{\omega l}\left(r\right)}\pm\sqrt{1-h}\right)=\nonumber\\
&=&\pm\frac{\omega r_{H}}{4c_{s}\delta r}\left(1\pm1+\mathcal{O}\left(\delta r\right)\right) \label{eq:log_drphi} \, .
\end{eqnarray}

When the constant $C$ takes positive values, the above expression describes solutions that peel off from the horizon and are relevant for the production of Hawking quanta, while for negative values of $C$ it describes horizon-crossing ingoing solutions.

From Equation~(\ref{eq:log_drphi}) it follows that the phase of each mode with positive $C$ diverges logarithmically, and on the two sides of the horizon we have two independent solutions. In the outer region these are outgoing modes, while in the inner region they peel off the horizon towards the singularity.
Considering the radial part of the modes, we have defined them either for $r>r_{H}$ or for  $0<r<r_{H}$. These functions can be analytically continued in the complex plane only assuming the presence of a branch cut.

In conclusion, the various modes are the  horizon-crossing modes $f_{\omega l m}^{\mathrm{HC}}$, the outgoing modes in the outer region $f_{\omega l m}^{\mathrm{ext}}$ and  their counterparts in the inner region $f_{\omega l m}^{\mathrm{int}}$.

\subsection{Far-Field Behavior}

At large radii the equation Equation~(\ref{eq:KG_norm}) is such that $h\approx 1$ and the mode-mixing potential $h\partial_{r} h/r$ is negligible, and the modes can be approximated with a linear combination of Bessel functions of the first and of the second kind. 

At infinity even the angular momentum potential can be neglected, and the properly normalized solutions take the form
\begin{eqnarray}
f_{\omega l}\left( r\right)\approx&\sqrt{\frac{\lambda}{4\pi\omega}}\frac{1}{r}e^{i\phi_{\omega l}^{\infty}}\left(\cos\Theta_{\omega l}e^{-i\omega r}+\sin\Theta_{\omega l}e^{i\omega r}\right) \, ,
\end{eqnarray}
where ingoing and outgoing modes are mixed at infinity, as expressed by the phase $\Theta_{\omega l}$ due to the potential encountered along the radial propagation.

In this region, apart from the overall phase, which is given by the limit $\lim_{r\rightarrow\infty}\int^{r}\frac{\sqrt{1-h}}{h}\mathrm{d}r$, and~by the conditions at the horizon, the radial part of the outgoing and the ingoing modes are one the conjugate of the other, with
\begin{eqnarray}
n_{\omega l}&=&\frac{\lambda}{4\pi\omega}\frac{1}{r^{2}}\left(1+\sin2\Theta_{\omega l}\cos\left(\omega r\right)\right) \, ,\\
\partial_{r}\phi_{\omega l}&=&\frac{C}{r^{2}n_{\omega l}}=\pm\frac{4\pi\omega^{2}}{\lambda c_{s}}\frac{r_{H}^{2}n_{\omega l}\left(r_{H}\right)}{\left(1+\sin2\Theta_{\omega l}\cos\left(\omega r\right)\right)} \, .
\end{eqnarray}  

\subsection{Mode Decomposition of the Scalar Field}

In conclusion, one obtains the mode decomposition of the scalar field, as a superposition of modes with fixed angular momentum and frequency
\begin{eqnarray}
\fl \quad
\theta_{1}\left(t,r,\Omega\right)&=&\int_{0}^{\infty}\mathrm{d}\omega\sum_{l=0}^{\infty}\sum_{m=-l}^{l}\left(f_{\omega lm}^{\mathrm{HC}}\left(t,r,\Omega\right)a_{\omega lm}^{\mathrm{HC}}+\overline{f_{\omega lm}^{\mathrm{HC}}}\left(t,r,\Omega\right)a_{\omega lm}^{\mathrm{HC}\dagger}\right)+\nonumber\\
\fl \quad
&&+\int_{0}^{\infty}\mathrm{d}\omega\sum_{l=0}^{\infty}\sum_{m=-l}^{l}\left(f_{\omega lm}^{\mathrm{ext}}\left(t,r,\Omega\right)a_{\omega lm}^{\mathrm{ext}}+\overline{f_{\omega lm}^{\mathrm{ext}}}\left(t,r,\Omega\right)a_{\omega lm}^{\mathrm{ext}\dagger}\right)+\nonumber\\
\fl \quad
&& +\int_{0}^{\infty}\mathrm{d}\omega\sum_{l=0}^{\infty}\sum_{m=-l}^{l}\left(f_{\omega lm}^{\mathrm{int}}\left(t,r,\Omega\right)a_{\omega lm}^{\mathrm{int}}+\overline{f_{\omega lm}^{\mathrm{int}}}\left(t,r,\Omega\right)a_{\omega lm}^{\mathrm{int}\dagger}\right) \, ,\\
\fl \quad
\rho_{1}\left(t,r,\Omega\right)&=&-\frac{1}{\lambda}\left(\partial_{t}-V\left(r\right)\partial_{r}\right)\theta_{1}\left(t,r,\Omega\right) \, .
\end{eqnarray}

Depending on the radius, only one set of modes between $f^{\mathrm{ext}}$ and $f^{\mathrm{int}}$ appears in the definition of the field.

The field operators follow the canonical commutation relations
\begin{eqnarray}
\left[\theta_{1}\left(t,r,\Omega\right),\theta_{1}\left(t,r^{\prime},\Omega^{\prime}\right)\right]&=&0 \, ,\\
\left[\theta_{1}\left(t,r,\Omega\right),\rho_{1}\left(t,r^{\prime},\Omega^{\prime}\right)\right]&=&\delta\left(\Omega,\Omega^{\prime}\right)\frac{\delta\left(r,r^{\prime}\right)}{rr^{\prime}}\, ,
\end{eqnarray}
and the ladder operators relative to the modes are defined to follow the usual commutation relations in the quasi-particle Fock space
\begin{eqnarray}
\left[a_{\omega lm}^{\mathrm{HC}},a_{\omega^{\prime}l^{\prime}m^{\prime}}^{\mathrm{HC}\dagger}\right]&=&\delta_{ll^{\prime}}\delta_{mm^{\prime}}\delta\left(\omega,\omega^{\prime}\right) \, ,\\
\left[a_{\omega lm}^{\mathrm{ext}},a_{\omega^{\prime}l^{\prime}m^{\prime}}^{\mathrm{ext}\dagger}\right]&=&\delta_{ll^{\prime}}\delta_{mm^{\prime}}\delta\left(\omega,\omega^{\prime}\right) \, ,\\
\left[a_{\omega lm}^{\mathrm{int}},a_{\omega^{\prime}l^{\prime}m^{\prime}}^{\mathrm{int}\dagger}\right]&=&\delta_{ll^{\prime}}\delta_{mm^{\prime}}\delta\left(\omega,\omega^{\prime}\right) \, ,
\end{eqnarray}
and all the other commutation relations between ladder operators vanish identically.

\section{Hawking Radiation}\label{sec:HRad}
%---------------------------------------------------------------------
We now discuss the creation of Hawking quanta. The presence of the horizon induces the creation of Hawking radiation, and the velocity profile defines the temperature of the canonical analogue black hole. Then we study the properties of the initial state, the Unruh vacuum state.

Until now we have focused on the region outside the horizon, where for every positive $\omega$ the modes $f_{\omega l m}^{\mathrm{ext}}$ and $f_{\omega l m}^{\mathrm{HC}}$---with positive norm with respect to the Klein--Gordon internal product---must have the time dependence $e^{-i \omega t}$ usual for stationary spacetimes, while in the inner region the time dependence is inverted for the modes with positive norm $f_{\omega l m}^{\mathrm{int}}$, because of the change of sign of the function $h\left(r\right)$.

Moreover, each of the modes peeling off from the horizon has a phase that diverges logarithmically, as can be deduced from Equation~(\ref{eq:log_drphi}), implying that on the two sides of the horizon the two independent sets of modes $f_{\omega l m}^{\mathrm{ext}}$ and $\overline{f_{\omega lm}^{\mathrm{int}}}$ can be analytically continued and put in superposition with each other.

The operation of mixing destruction operators of the modes defined in the
outer region with creation operators of the modes defined in the inner
region is described by a Bogoliubov transformation.

In particular we are interested in the mixing between modes at the horizon with those associated with a vacuum state for static observers at past null infinity.

Near the horizon, at the two sides, the modes are (dropping the eigenvalues from the notation)
\begin{eqnarray}
f^{\mathrm{ext}}=&\sqrt{n^{\mathrm{ext}}}e^{-i\omega t}e^{i\frac{\omega r_{H}}{2c_{s}}\ln\frac{\left|\delta r\right|}{r_{H}}}\Theta\left(\delta r\right)\, ,\\
\overline{f^{\mathrm{int}}}=&\sqrt{n^{\mathrm{int}}}e^{-i\omega t}e^{i\frac{\omega r_{H}}{2c_{s}}\ln\frac{\left|\delta r\right|}{r_{H}}}\Theta\left(-\delta r\right) \, .
\end{eqnarray}

Linear combinations of these modes define new solutions, in particular we are interested in the analytic continuations of the radial part on the real $r$-axis. One solution can be extended in the upper half-plane---with a branch cut in the lower, and one in the lower half-plane---with a branch cut in the~upper.
\begin{eqnarray}
f_{+}&=&\alpha_{1}f^{\mathrm{ext}}+\beta_{1}\overline{f^{\mathrm{int}}} \, ,\\
f_{-}&=&\beta_{2}f^{\mathrm{ext}}+\alpha_{2}\overline{f^{\mathrm{int}}}\, .
\end{eqnarray}

Due to the logarithmic term in the phase of the functions, passing from one side of the horizon to the other, the analytic continuations gain a phase of $\pm\pi\omega r_{H}/2c_{s}$, and therefore we can write
\begin{eqnarray}
\alpha_{1}\sqrt{n^{\mathrm{ext}}}&=&e^{\frac{\pi\omega r_{H}}{2c_{s}}}\beta_{1}\sqrt{n^{\mathrm{int}}} \, ,\\
\beta_{2}\sqrt{n^{\mathrm{ext}}}&=&e^{-\frac{\pi\omega r_{H}}{2c_{s}}}\alpha_{2}\sqrt{n^{\mathrm{int}}} \, .
\end{eqnarray}

The coefficients $\alpha$ and $\beta$ of the Bogoliubov transformation can be found imposing that the field can be rewritten in terms of these new modes and imposing the canonical commutation relations for the new ladder operators:
\begin{eqnarray}
\qquad \quad f_{+}a_{+}+f_{-}a_{-}^{\dagger}&=&f^{\mathrm{ext}}a^{\mathrm{ext}}+\overline{f^{\mathrm{int}}}a^{\mathrm{int}\dagger} \, ,\\
\left[a_{+},a_{+}^{\dagger}\right]=\left[a_{-},a_{-}^{\dagger}\right]&=&\left[a^{\mathrm{ext}},a^{\mathrm{ext}\dagger}\right]=\left[a^{\mathrm{int}},a^{\mathrm{int}\dagger}\right] \, .
\end{eqnarray}

Therefore, it is found that apart from irrelevant overall phases, the Bogoliubov transformation is given by
\begin{eqnarray}
\left|\alpha_{1}\right|^{2}=\left|\alpha_{2}\right|^{2}&=&\frac{e^{\frac{\pi\omega r_{H}}{c_{s}}}}{e^{\frac{\pi\omega r_{H}}{c_{s}}}-1} \label{alpha}\, ,\\
\left|\beta_{1}\right|^{2}=\left|\beta_{2}\right|^{2}&=&\frac{1}{e^{\frac{\pi\omega r_{H}}{c_{s}}}-1}\label{beta} \, .
\end{eqnarray}

The Unruh vacuum $\left|\emptyset\right\rangle$ is defined such that
\begin{eqnarray}
a_{+}\left|\emptyset\right\rangle &=&0 \, , \label{eq:UV1}\\
a_{-}\left|\emptyset\right\rangle &=&0\, ,\label{eq:UV2}\\
a^{\mathrm{HC}}\left|\emptyset\right\rangle &=&0 \, ,\label{eq:UV3}
\end{eqnarray}

So, Equations~(\ref{alpha}) and (\ref{beta}) imply that an observer outside the horizon will observe a thermal distribution of radiation coming from the black hole horizon. Indeed, through the Bogoliubov transformation, the occupation of the modes $f^{\mathrm{ext}}$ is given by
\begin{eqnarray}
\left\langle a_{\omega lm}^{\mathrm{ext}\dagger}a_{\omega lm}^{\mathrm{ext}}\right\rangle =&\frac{1}{e^{\frac{\pi\omega r_{H}}{c_{s}}}-1}\delta_{ll^{\prime}}\delta_{mm^{\prime}}\delta\left(\omega,\omega^{\prime}\right) \, . \label{eq:occup_num}
\end{eqnarray}

This occupation number corresponds to the thermal Bose--Einstein distribution associated with the~temperature
\begin{eqnarray}
T &=&\frac{c_{s}}{\pi r_{H}}=\frac{c_{s}\kappa}{2\pi} \, ,\label{eq:temperature}\\
\kappa &=& \frac{2}{r_{H}}\, ,
\end{eqnarray}
where $\kappa$ is the surface gravity of the canonical analogue black hole, which is half the radial derivative of $h$ evaluated at the horizon
\begin{eqnarray}
\kappa=\left[\frac{1}{2}\partial_{r}h\right]_{r_{H}} \, .
\end{eqnarray}

\section{Back-Reaction}\label{sec:BReac}
%---------------------------------------------------------------------
When the Gross--Pitaevskii equation is modified by introducing the terms of anomalous mass and density, it becomes possible to study the back-reaction that the analogous scalar field induces on the condensate wavefunction.

The study of the back-reaction in analogue systems enables the address of various issues, such as the problem of the evaporation of black holes and in general the dynamics of analogue horizon. To study the dynamics of the horizon it is necessary to express the anomalous mass and density as correlation functions of the analogous scalar field and its conjugate. Although the anomalous density is strictly real, the anomalous mass contains a real and an imaginary part.

It is particularly convenient to study these quantities in dimensionless terms:
\begin{eqnarray}
n &=&\frac{\left\langle \delta\phi^{\dagger}\delta\phi\right\rangle }{\overline{\left\langle \phi_{0}\right\rangle }\left\langle \phi_{0}\right\rangle }=\frac{1}{4}\left\langle \frac{\rho_{1}}{\rho_{0}}\frac{\rho_{1}}{\rho_{0}}\right\rangle+\left\langle \theta_{1}\theta_{1}\right\rangle \label{eq:n_AD} \, ,\\
m_{R} &=&\frac{1}{2}\left(\frac{\left\langle \delta\phi\delta\phi\right\rangle }{\left\langle \phi_{0}\right\rangle \left\langle \phi_{0}\right\rangle }+\frac{\left\langle \delta\phi^{\dagger}\delta\phi^{\dagger}\right\rangle }{\overline{\left\langle \phi_{0}\right\rangle }\overline{\left\langle \phi_{0}\right\rangle }}\right)=\frac{1}{4}\left\langle \frac{\rho_{1}}{\rho_{0}}\frac{\rho_{1}}{\rho_{0}}\right\rangle -\left\langle \theta_{1}\theta_{1}\right\rangle  \label{eq:mR_AMR} \, ,\\
m_{I} &=&\frac{1}{2i}\left(\frac{\left\langle \delta\phi\delta\phi\right\rangle }{\left\langle \phi_{0}\right\rangle \left\langle \phi_{0}\right\rangle }-\frac{\left\langle \delta\phi^{\dagger}\delta\phi^{\dagger}\right\rangle }{\overline{\left\langle \phi_{0}\right\rangle }\overline{\left\langle \phi_{0}\right\rangle }}\right)=\frac{1}{2}\left\langle \left\{\frac{\rho_{1}}{\rho_{0}},\theta_{1}\right\} \right\rangle =\nonumber\\
&=& -\frac{1}{2}\frac{1}{\lambda\left\langle \rho_{0}\right\rangle }\left(\partial_{t}+\frac{\nabla\left\langle\theta_{0}\right\rangle}{m}\nabla\right)\left\langle \theta_{1}\theta_{1}\right\rangle  \, .\label{eq:mI_AMI}
\end{eqnarray}

We observe that in Equation~(\ref{eq:mI_AMI}), in the hydrodynamic limit, the imaginary part of the anomalous density is the derivative along the flow of the $\left\langle \theta_{1}\theta_{1} \right\rangle $ correlation function, the vacuum polarization. Please note that in Equation~(\ref{eq:n_AD}) we have regularized the definition of anomalous density, thus eliminating a $\delta^{3}\left(0\right)$ term. Formally this is done defining the correlation function as the limit for vanishing distance of the 2-point correlation function.

To describe the mean-field dynamics with the inclusion of the anomalous terms we modify the Gross--Pitaveskii equation as in Equation~(\ref{eq:GP+BR}) and apply the Madelung representation.~(Let us stress that while these are definitely not the Einstein equations, they can nonetheless be cast in the form of a modified Poisson equation~\cite{Sindoni:2009fc}.)

We study separately the square norm of the wavefunction---the atom density of the condensate---and the phase:
\begin{eqnarray}
\fl\quad
\partial_{t}\left\langle \rho\right\rangle &=&-\frac{1}{m}\nabla\left(\left\langle \rho\right\rangle \left(\nabla\left\langle \theta\right\rangle \right)\right)+2\lambda\left\langle \rho\right\rangle ^{2}m_{I} \label{eq:MF_rho} \, ,\\
\fl\quad
\partial_{t}\left\langle \theta\right\rangle &=&\frac{1}{2m}\left\langle \rho\right\rangle ^{-1/2}\nabla^{2}\left\langle \rho\right\rangle ^{1/2}-\frac{1}{2m}\left(\nabla\left\langle \theta\right\rangle \right)\left(\nabla\left\langle \theta\right\rangle \right)-\lambda\left\langle \rho\right\rangle \left(1+2n+m_{R}\right)-V_{\mathrm{ext}} \label{eq:MF_theta} \, .
\end{eqnarray}

The equation for the dynamics of the atom density function is affected by the imaginary part of the anomalous mass, while the equation for the dynamics of the phase of the condensate wavefunction is modified by a combination of anomalous density and real part of the anomalous mass.

We study the regime in which these classical perturbations to the condensate wavefunction are small deviations from the stationary solution of the Gross--Pitaevkii equation that do not break the hydrodynamic approximation.

Therefore, we neglect the terms that are suppressed by the healing length scale $\xi=1/\sqrt{\lambda\left\langle \rho\right\rangle m}$, because they are only relevant at a high energy scale where the interactions between the individual atoms become dominant in determining the phenomenology. In Equations~(\ref{eq:MF_rho}) and~(\ref{eq:MF_theta}) we neglect the contributions to the anomalous density and to the real part of the anomalous mass which goes as $\left\langle\rho_{1}\rho_{1}\right\rangle$. Since the definition of the conjugate field $\rho_{1}$ depends on the healing length scale as 
\begin{eqnarray}
\frac{\rho_{1}}{\left\langle \rho_{0}\right\rangle }=&-\frac{1}{\lambda\left\langle \rho_{0}\right\rangle }\left(\partial_{t}+\frac{\nabla\left\langle \theta_{0}\right\rangle }{m}\nabla\right)\theta_{1}=-\frac{\xi}{c_{s}}\left(\partial_{t}+\frac{\nabla\left\langle \theta_{0}\right\rangle }{m}\nabla\right)\theta_{1}\, ,
\end{eqnarray}
the anomalous density in Equation~(\ref{eq:n_AD}) and the real part of the anomalous mass in Equation~(\ref{eq:mR_AMR}) can be approximated to the contributions of the correlation function $\left\langle \theta_{1}\theta_{1}\right\rangle $:
\begin{eqnarray}
n \approx -m_{R} \approx \left\langle \theta_{1}\theta_{1}\right\rangle \, .
\end{eqnarray}

Under these assumptions, from Equations~(\ref{eq:MF_rho}) and~(\ref{eq:MF_theta}) we proceed linearizing the equations for the mean-field for small perturbations---of the same order of magnitude of the anomalous terms---of the solution of the Gross--Pitaevskii equation, getting
\begin{eqnarray}
\left\langle \rho\right\rangle &=&\left\langle \rho_{0}\right\rangle +\left\langle \delta\rho\right\rangle \, ,\label{eq:variation_density}\\
\left\langle \theta\right\rangle &=&\left\langle \theta_{0}\right\rangle +\left\langle \delta\theta\right\rangle \, ,\label{eq:variation_phase} 
\end{eqnarray}
\begin{eqnarray}
\left(\partial_{t}+\frac{\nabla\left\langle \theta_{0}\right\rangle }{m}\nabla\right)\left(\frac{\left\langle \delta\rho\right\rangle }{\left\langle \rho_{0}\right\rangle }+\left\langle \theta_{1}\theta_{1}\right\rangle \right)=&-\frac{1}{m\left\langle \rho_{0}\right\rangle }\nabla\left(\left\langle \rho_{0}\right\rangle \nabla\left\langle \delta\theta\right\rangle \right) \label{eq:dyndrho} \, ,
\end{eqnarray}
\begin{eqnarray}
\left(\partial_{t}+\frac{\nabla\left\langle \theta_{0}\right\rangle }{m}\nabla\right)\left\langle \delta\theta\right\rangle =&-\lambda\left\langle \rho_{0}\right\rangle \left(\frac{\left\langle \delta\rho\right\rangle }{\left\langle \rho_{0}\right\rangle }+\left\langle \theta_{1}\theta_{1}\right\rangle \right) \label{eq:dyndtheta} \, .
\end{eqnarray}

We can see that the classical fluctuation of the phase of the condensate $\left\langle \delta\theta\right\rangle $ evolves like a Klein--Gordon mode, with conjugate momentum $\frac{\left\langle \delta\rho\right\rangle }{\left\langle \rho_{0}\right\rangle }+\left\langle \theta_{1}\theta_{1}\right\rangle $.

These equations allow study of the dynamics of the classical perturbation, and are differential equations that can be solved after the source term $\left\langle \theta_{1}\theta_{1}\right\rangle $ and the initial conditions for $\left\langle \delta\rho\right\rangle $ and $\left\langle \delta\theta\right\rangle $ are provided. Unless broken by the initial conditions, the evolution of the classical perturbation will preserve the spherical symmetry and the time translation symmetry.

\subsection{Dynamics of the Horizon}
The production of Hawking radiation emitted by a black hole formed by gravitational collapse is a process that transfers energy towards the future null infinity.
The outgoing flux of Hawking quanta induces a loss of mass of the black hole, whose apparent horizon gradually shrinks, until its eventual evaporation. In this non-stationary process, the gravitational collapse triggers the production of Hawking radiation quanta, consequently exerting a back-reaction on the classical metric tensor.

In the case of the canonical analogue black hole, we can study the dynamics of the horizon assuming the condensate to be initially in the stationary condensate configuration described in Section~\ref{sec:CABH}. Then, turning on the production of quasi-particles of the field $\theta_{1}$, we observe a back-reaction on the wavefunction. 

In general, with spherical symmetry, the horizon radius $r_{H}$ is defined as the radius at which the radial velocity of the condensate equals the speed of sound. In the unperturbed case we have
\begin{eqnarray}
0=&\left[c_{s}^{2}-V_{0}^{2}\right]_{r_{H}} \, .\label{eq:position_horizon}
\end{eqnarray}

If the condensate is in a state which is a small spherically symmetric perturbation of the stationary solution of the Gross--Pitaevskii equation, the density and the phase of the condensate are modified as described in Equations~(\ref{eq:variation_density}) and (\ref{eq:variation_phase}).~(Notice that we are assuming that both the perturbed condensate wavefunction and the quasi-particles are in a globally spherically symmetric state, in which modes of opposite angular momenta have the same amplitude. The Hamiltonian driving the evolution ensures the conservation of the total angular momentum which is initially vanishing.) Consequently, the velocity of the condensate and the speed of sound change and, under this variation, Equation~(\ref{eq:position_horizon}) is satisfied at a different radius $r_{H}+\delta r_{H}$, where  $r_{H}$ is the unperturbed horizon radius:
\begin{eqnarray}
V_{0} &\rightarrow& V=V_{0}+\delta V=-\frac{\partial_{r}\left\langle \theta_{0}\right\rangle }{m}-\frac{\partial_{r}\left\langle \delta\theta\right\rangle }{m} \, , \\
c_{s}^{2}&\rightarrow & c_{s}^{2}\left(1+\frac{\left\langle \delta\rho\right\rangle }{\left\langle \rho_{0}\right\rangle }\right) \, , \\
r_{H}&\rightarrow&r_{H}+\delta r_{H}=r_{H}-\left[\frac{\delta\left(V^{2}/c_{s}^{2}\right)}{\partial_{r}\left(V_{0}^{2}/c_{s}^{2}\right)}\right]_{r_{H}} \label{eq:deltarh} \, .
\end{eqnarray}

The denominator in Equation~(\ref{eq:deltarh}) is different for each spherically symmetric system and is proportional to the surface gravity of the horizon, i.e., to the temperature of the black hole in Equation~(\ref{eq:temperature})
\begin{eqnarray}
\left[-\partial_{r}\left(V^{2}/c_{s}^{2}\right)\right]_{r_{H}}=&2\kappa>0 \, .
\end{eqnarray}

The radius of the perturbed horizon is smaller or larger than $r_{H}$ depending on the sign of the variation $\delta\left(V^{2}/c_{s}^{2}\right)$.

For the canonical analogue black hole, we substitute the expression of $V_{0}$ of Equation~(\ref{eq:definition_V0}) in Equation~(\ref{eq:deltarh}) and we get
\begin{eqnarray}
\delta r_{H}&=&-\frac{r_{H}}{4}\left[\frac{\left\langle \delta\rho\right\rangle }{\left\langle \rho_{0}\right\rangle }-2\frac{\delta V}{c_{s}}\right]_{r_{H}} \, .
\end{eqnarray}

Deriving the variation of the horizon radius with respect to time we observe which quantities determine the dynamics of $\delta r_{H}$
\begin{eqnarray}
\partial_{t}\delta r_{H}=&-\left[\frac{\partial_{t}\delta\left(V^{2}/c_{s}^{2}\right)}{\partial_{r}\left(V^{2}/c_{s}^{2}\right)}\right]_{r_{H}}-\delta r_{H}\left[\partial_{t}\ln\partial_{r}\left(V^{2}/c_{s}^{2}\right)\right]_{r_{H}} \, .
\end{eqnarray}

For a stationary spacetime the second term vanishes. 

We consider the case of the canonical analogue black hole substituting from Equations~(\ref{eq:dyndrho}) and~(\ref{eq:dyndtheta}) the expressions for $\delta V$ and $\left \langle \delta \rho \right \rangle$:
\begin{eqnarray}
\partial_{t}\delta r_{H}&=&\frac{r_{H}c_{s}}{4}\left[\partial_{r}\left(\frac{\left\langle \delta\rho\right\rangle }{\left\langle \rho_{0}\right\rangle }+\left\langle \theta_{1}\theta_{1}\right\rangle +\frac{\delta V}{c_{s}}\right)\right]_{r_{H}}+\left[
%\frac{r_{H}}{4}\partial_{t}\left\langle \theta_{1}\theta_{1}\right\rangle 
-\frac{3}{2}\delta V\right]_{r_{H}} \label{eq:drho_dyn} \, .
%\\
%&=&\left[-\delta V+\frac{r_{H}}{4}\partial_{t}\left(\left\langle \theta_{1}\theta_{1}\right\rangle +\frac{\delta V}{c_{s}}\right)\right]
\end{eqnarray}

The source term  on the RHS is a quantity that depends on the quantum state of the field $\theta_{1}$---through the correlation function $\left\langle \theta_{1}\theta_{1}\right\rangle $---and on the state of the classical perturbation. It does not have definite sign: different initial conditions for the classical perturbation determine different regimes of the black hole,  i.e.,~expansion or contraction. This is expected, as Equation~(\ref{eq:drho_dyn}) applies to every possible regime of the black hole.

Nonetheless, such different initial conditions for the classical perturbation do not {affect}, the~contribution from the quantum field $\theta_{1}$, which remains always the same. In the hierarchy of the equations, the analogue scalar field affects the perturbation of the analogue metric, not \emph{vice versa}. A~self-consistent semiclassical approach would require including the perturbation of the wavefunction in the Bogoliubov--de Gennes equations, explicitly or by iteratively recalculating $\left\langle \theta_{1}\theta_{1}\right\rangle $ and the pair $\left( \left \langle \delta\rho \right \rangle, \left \langle \delta \theta \right \rangle \right)$.

The effect of the analogue Hawking radiation is understood when $\left\langle \theta_{1}\theta_{1}\right\rangle $ is the contribution leading the phenomenology. We therefore assume the perturbation terms $\left\langle \delta\theta\right\rangle $ and $\left\langle \delta\rho\right\rangle $ and their radial derivatives to be negligible. In the analogy between the canonical analogue black hole and the gravitational black hole, this corresponds to assuming the initial perturbation of the metric to be negligible, and then to excite it through the presence of Hawking radiation. This reflects the evolution of a gravitational black hole, where the free-falling matter induces the formation of the horizon, which~triggers the production of Hawking quanta slowly driving the spacetime far from the stationary configuration.

Therefore, we consider Equation~(\ref{eq:drho_dyn}) and assume the perturbation of the wavefunction to be null and the corresponding terms to vanish. Moreover, we can assume the time derivative of the Hawking radiation term to be negligible. The Hawking quanta are produced in pairs of equal frequency, and~in the correlation functions their time dependent phases cancel each other out. The term $\partial_{t}\left\langle \theta_{1}\theta_{1}\right\rangle $ is therefore suppressed.

In this regime, we are left with
\begin{eqnarray}
\partial_{t}\delta r_{H}&\approx &\frac{r_{H}c_{s}}{4}\left[\partial_{r}\left\langle \theta_{1}\theta_{1}\right\rangle \right]_{r_{H}} \, .
\end{eqnarray}

The quantity $\left\langle \theta_{1}\theta_{1}\right\rangle$  is of paramount interest in the investigation of the back-reaction that quantum fields exert on curved geometries. Together with the stress-energy tensor, the vacuum polarization is a quantity that in the literature is studied in association with the production of Hawking quanta in gravitational systems~\cite{Page:1982fm,Anderson:1990jh}.

As described in~\cite{Christensen:1976vb,Candelas:1980zt} this function is obtained through the coincidence limit of the Green function of the Klein--Gordon operator, i.e., the limit for $x^{\prime}\rightarrow x$ and $t^{\prime}\rightarrow t$ of $G\left(t,x;t^{\prime},x^{\prime}\right)$. Different boundary conditions determine different expressions for $\left\langle \theta_{1}\theta_{1}\right\rangle$: the occupation numbers of the modes of the field, which are associated at the horizon with the Hawking radiation and with the horizon-crossing quanta, determine the behavior of this function. 
In particular, the occupation numbers are zero for the Boulware vacuum; are thermal for the outgoing modes and null for the ingoing modes for the Unruh vacuum (as seen in Equations~(\ref{eq:UV1})--(\ref{eq:UV3})); and are all thermal for the Hartle--Hawking vacuum. To obtain the expression for $\left\langle \theta_{1}\theta_{1}\right\rangle$ in the various cases one must subtract different renormalization counterterms. Renormalizing is strictly necessary as the Green function generally includes divergent terms. In the coincidence limit already in the 3+1 flat Minkowski spacetime it is found that
\begin{eqnarray}
G\left(t+\epsilon,x;t,x\right)\propto	-\frac{1}{\epsilon^{2}} \, .
\end{eqnarray}

The most commonly used renormalization scheme is based on the point splitting regularization method~\cite{Schwinger:1951nm,DeWitt:1975ys,Christensen:1976vb} which removes the divergence by splitting the point in which the Green functions is evaluated in two nearby by points characterized by their geodesic distance, so regularizing the vacuum polarization as measured along geodesic trajectories. 

For Hadamard states~\cite{Birrell:1982ix} the resulting divergent structure in the coincidence limit is universal in curved spacetimes: one gets the above-mentioned divergent terms together with other logarithmically divergent terms typical of the Hadamard structure.  The universality of such ultraviolet divergences (of the same functional form of those obtainable in Minkowski spacetime), allows safe discarding of them.~(Observing that the renormalized stress-energy tensor in Minkowski (flat) spacetime in vacuum should be zero, then the ultraviolet divergence found in quantum field theory can be deemed unphysical and should be discarded. Hence, when the very same contribution is found in curved spacetimes (which are locally Minkowski) it should be subtracted applying the same renormalization scheme.) However, other irregular behaviors may nonetheless arise from the peculiarities of the curved geometry and the vacuum state. In particular, in the presence of a horizon, the Boulware vacuum gives a divergent vacuum polarization due to the vanishing of the time-time element of the metric $h$. As argued by Candelas, in proximity of the horizon the Green function is divergent as $G=-1/{h \epsilon^2}$~\cite{Candelas:1980zt}. Also, let us mention that while the above-mentioned regularization schemes have been mostly applied in Ricci flat spacetimes, in non-Ricci flat spacetimes such as ours, they will generically include an extra contribution, which however we can expect to provide at the horizon at most a prefactor of order unity, which can be therefore neglected for our considerations.

When we consider the Bogoliubov transformations, the Green function depends on the occupation numbers of the quasi-particle states. For states different from the Boulware vacuum, in the presence of non-null occupation numbers, the vacuum polarization is given by the limit
\begin{eqnarray}
\left\langle \theta_{1}\theta_{1}\right\rangle \propto\frac{1}{h}\lim_{\epsilon\rightarrow0^{*}}\int_{0}^{\infty}\mathrm{d}\omega \, e^{-i\omega\epsilon} \omega \left(1+2n_{\omega}\right) \, , \label{eq:integral_expression}
\end{eqnarray}
where the first contribution is fixed for every state.

To appropriately regularize the contribution of the Hawking radiation, it is, therefore, necessary to subtract the quadratic divergence that defines the Boulware vacuum and consider the occupation numbers $n_{\omega}$. The Unruh vacuum is associated with the occupation numbers of Equation~(\ref{eq:occup_num}), for~which the integral expression in Equation~(\ref{eq:integral_expression}) gives
\begin{eqnarray}
\frac{2}{h}\int_{0}^{\infty}\mathrm{d}\omega\omega n_{\omega}=&\frac{2}{h}\int_{0}^{\infty}\mathrm{d}\omega\frac{\omega}{e^{\frac{2\pi\omega}{c_{s}\kappa}}-1}=\frac{c_{s}^{2}\kappa^{2}}{6h} \, .
\end{eqnarray}

This quantity is divergent at the horizon, but is regularized by subtraction of the second term in the series expansion of the geodesic distance $\frac{c_{s}^{2}}{6h}\left(\frac{h^{\prime}}{2}\right)^{2}$, which removes the divergence due to $h$ in the limit $r\rightarrow r_{H}$.

Reintroducing the proper numerical factors, we therefore obtain that in proximity of the horizon the leading effect of the Hawking radiation induces the perturbative expression
\begin{eqnarray}
\left\langle \theta_{1}\theta_{1}\right\rangle &\approx&\frac{\lambda}{c_{s}}\frac{1}{48\pi^{2}h}\left(\kappa^{2}-\left(\frac{h^{\prime}\left(r\right)}{2}\right)^{2}\right)=\label{eq:general_theta2}\\
&=&\frac{\lambda}{c_{s}}\frac{1}{96\pi^{2}}\left(-h^{\left(2\right)}\left(r_{H}\right)-\frac{1}{2}h^{\left(3\right)}\left(r_{H}\right)\delta r+\mathcal{O}\left(\delta r^{2}\right)\right)=\\
&=&\frac{\lambda}{c_{s}}\frac{1}{96\pi^{2}}\frac{20}{r_{H}^{2}}\left(1-3\frac{\delta r}{r_{H}}+\mathcal{O}\left(\delta r^{2}\right)\right) \label{eq:CABH_theta2} \, .
\end{eqnarray}
	
This approximation does not hold for every radius, e.g., it includes not integrable divergences in the origin, but is such that it is consistent with the expected behavior of the Rindler spacetime---the near-horizon region~\cite{Boulware:1974dm,Frolov:1998wf}. In Equation~(\ref{eq:CABH_theta2}) we retain the term of order $\delta r$, as we are interested in the radial derivative of this expression.

Please note that the adimensionality of the field is a feature that differs from the usual notation, as in 3+1 quantum field theory the Klein--Gordon scalar field has generally the dimension of an inverse length. In Equation~(\ref{eq:general_theta2}) we introduced the length scale $\sqrt{\lambda/c_{s}}$, which is the proper quantity needed to replicate the usual relations
\begin{eqnarray}
&&S=-\int\mathrm{d}^{4}x\frac{1}{2}\sqrt{g}g^{\mu\nu}\left(\partial_{\mu}\theta\right)\left(\partial_{\nu}\theta\right) \, ,\\
&&\pi = -\frac{\delta\mathcal{L}}{\delta\partial_{0}\theta}=\sqrt{g}g^{0\mu}\left(\partial_{\mu}\theta\right) \, ,\\
&&\left[\theta\left(x\right),\pi\left(y\right)\right]=-i\delta^{3}\left(x,y\right) \, .
\end{eqnarray}

Considering the definition of $\rho_{1}$ given in Equation~(\ref{eq:negl_QP}), a straightforward calculation gives the length scale required.

Therefore, we obtain the estimate for the time derivative of the horizon radius, and from the expression of the surface gravity we can also make an estimate of its rate of change
\begin{eqnarray}
\partial_{t}\delta r_{H}&\approx&\frac{r_{H}c_{s}}{4}\partial_{r}\left\langle \theta_{1}\theta_{1}\right\rangle =\\
&=&-\frac{5}{32\pi^{2}}\frac{\lambda}{r_{H}^{2}}= \label{eq:dtrh}\\
&=&-\frac{5}{32\pi^{2}}\frac{c_{s}}{\xi r_{H}^{2}\left\langle\rho\right\rangle} \, ,\\
\frac{1}{c_{s}}\frac{\dot{\kappa}}{\kappa^{2}}&\approx &\frac{5}{64\pi^{2}}\frac{1}{\xi r_{H}^{2}\left\langle\rho\right\rangle}\ll 1 \label{eq:k_dot} \, .
\end{eqnarray}

These derivatives are very small: while typically the atom separation---equal to $\rho_{0}^{-1/3}$---is of an order comparable to the healing length, the radius of the horizon can safely be assumed being much larger than both.

Given that the rate of change of the surface gravity is very small, the system can be assumed to evolve adiabatically along the evolution: the hydrodynamic approximation is broken before the rate of change of the curvature in Equation~(\ref{eq:k_dot}) becomes comparable to 1.

We can therefore take Equation~(\ref{eq:dtrh}) to provide a rough approximate description of the evolution of the black hole, promoting $r_{H}$ on the RHS to be the dynamical horizon radius. We obtain that the expected lifetime of the black hole is long and proportional to the inverse of the interaction coupling $\lambda$
\begin{eqnarray}
r_{H}\left(t\right)&=&\left(r_{H}^{3}\left(t_{0}\right)-\frac{15\lambda}{32\pi^{2}}t\right)^{1/3} \, ,\\
t_{\mathrm{fin}}&=&\frac{32\pi^{2}r_{H}^{3}\left(t_{0}\right)}{15\lambda}=\frac{8\pi}{5}N_{BH}t_{\mathrm{Healing}} \label{eq:t_fin}\, ,
\end{eqnarray}
where $N_{BH}$ is the initial average number of condensate atoms in the region within the horizon 
\begin{eqnarray}
N_{BH}&=&\frac{4\pi}{3} \left\langle\rho_{0}\right\rangle r^3_{H}\left(t_{0}\right)\, ,
\end{eqnarray}
and $t_{\mathrm{Healing}}=\xi/c_{s}$.
Let us note that the cubic dependence on $r_{H}$ in Equation~(\ref{eq:t_fin}) resembles that of a Schwarzschild black hole, and follows from the proportionality between $\delta \dot r_H$ and $r_{H}^{-2}$. To make this prediction it is necessary to assume the regime in which the Hawking radiation, i.e., the contribution of the Bogoliubov quasi-particles obtained from the solution of the Bogoliubov--de Gennes equations, is~dominant over the perturbation obtained by back-reaction in the modified Gross--Pitaevskii equation. Also, it must be kept in mind that the hydrodynamic approximation on which analogue gravity rests would break down when the black hole radius starts to be comparable with the healing length.
Nonetheless, Hawking radiation back-reaction acts over several healing times, see Equation~(\ref{eq:t_fin}), hence acoustic black holes several healing lengths in size would take a very long time to reach this point. Moreover, in realistic experiments, the end of the Hawking radiation regime is usually induced by hydrodynamic considerations (stability of the acoustic geometry) rather than by the back-reaction of the Hawking flux~\cite{kolobov2019spontaneous}.

We briefly point out that another regime which could be of interest is that of an analogue black hole in equilibrium with the analogue scalar field, where the back-reaction is not neglected but the classical perturbation is stationary. This would be the analogue of the Hartle--Hawking vacuum state for our system. In this case, a solution can be found without changing the speed profile, i.e., with~$\delta V\approx0$, but with a perturbation of the number density $\left\langle\delta\rho\right\rangle$ such that it counterbalances $\left \langle \theta_{1}\theta_{1}\right\rangle$:
\begin{eqnarray}
0&=&\frac{\left\langle \delta\rho\right\rangle }{\left\langle \rho_{0}\right\rangle }+\left\langle \theta_{1}\theta_{1}\right\rangle \, ,\\
0&=&\left\langle \delta\theta\right\rangle \, .
\end{eqnarray}

In this case, the horizon for the stationary configuration is found at a radius larger than $r_{H}$, but this solution will be driven out of equilibrium by the terms of order $\xi^{2}$ and the non-linearities previously neglected. In this case, the density–density correlation function differs from the unperturbed case by a negative quantity on the diagonal (when the density operators are evaluated at the same position):
\begin{eqnarray}
\left\langle \rho\rho\right\rangle -\left\langle \rho_{0}\right\rangle \left\langle \rho_{0}\right\rangle \approx&2\left\langle \rho_{0}\right\rangle \left\langle \delta\rho\right\rangle =-2\left\langle \rho_{0}\right\rangle ^{2}\left\langle \theta_{1}\theta_{1}\right\rangle \, .
\end{eqnarray}

If in a realization of the system the evolution is kept under control allowing only adiabatic transformations between near equilibrium configurations, it would be reasonable to assume these initial conditions for the system; but this equilibrium configuration would not be analogous to the Unruh vacuum.

\subsection{Dynamics of the Number of Atoms in the Condensate}
From Equation~(\ref{eq:dyndrho}) we observe that the dynamics of the perturbation of the condensate affects not only the position of the horizon, but also the global properties of the system, such as the number of atoms in the condensate state.

Let us consider the dynamics of $\left\langle \delta\rho\right\rangle$. The integral of this quantity in space gives the rate at which the atoms leave the condensate and move to the excited part. It is a global process which can already be described at the level of the Bogoliubov--de Gennes equations, considering the derivative with respect to time of the total number of atoms in the condensate 1-particle state, and assuming the stationarity of the unperturbed condensate wavefunction: 
\begin{eqnarray}
\partial_{t}N=\cancel{\partial_{t}N_{\mathrm{TOT}}}-\partial_{t}\int\mathrm{d}x\left\langle \delta\phi^{\dagger}\delta\phi\right\rangle \, .
\end{eqnarray}

The same derivative can be expressed as the derivative of the integral of the number density, i.e., of the squared norm of the condensate wavefunction. Since the unperturbed configuration is assumed stationary, the only contribution is made by the classical perturbation, i.e., by the back-reaction of the analogue quantum field. From Equation~(\ref{eq:dyndrho}) we get
\begin{eqnarray}
\fl\quad
\partial_{t}N&=&\partial_{t}\int\mathrm{d}x\left\langle \delta\rho\right\rangle =\\
\fl\quad
&=&-\int\mathrm{d}x\left\langle \rho_{0}\right\rangle \left(\frac{\nabla\left\langle \theta_{0}\right\rangle }{m}\right)\nabla\left\langle \theta_{1}\theta_{1}\right\rangle -\frac{1}{m}\cancel{\int\mathrm{d}x\nabla\left(\left\langle \delta\rho\right\rangle \nabla\left\langle \theta_{0}\right\rangle +\left\langle \rho_{0}\right\rangle \nabla\left\langle \delta\theta\right\rangle \right)}= \\
\fl\quad
&=&4\pi c_{s}r_{H}^{2}\left\langle \rho_{0}\right\rangle \int_{0}^{\infty}\mathrm{d}r\partial_{r}\left\langle \theta_{1}\theta_{1}\right\rangle = \label{eq:int_drVP}\\
\fl\quad
&=& -3N_{BH} T \left[\left\langle \theta_{1}\theta_{1}\right\rangle \right]_{0} .
\end{eqnarray}

We observe that the time derivative of the number of atoms in the condensate---in the entire occupied volume---depends on three factors: the number of condensate atoms in the region within the horizon $N_{BH}$ ; the Hawking temperature of the black hole, as defined in Equation~(\ref{eq:temperature}), that goes with the inverse of the horizon radius (plus subleading corrections describing the imperfection in the adiabaticity of the evolution); the value that the vacuum polarization $\left \langle \theta_{1}\theta_{1}\right\rangle$ assumes at the origin, which is effectively a dimensionless structure factor depending on the velocity profile. The vacuum polarization requires a proper renormalization, and it is generally difficult to extend it down to the singularity in the origin, not only when studying the canonical analogue black hole but also, e.g., for~the Schwarzschild black hole~\cite{Candelas:1985ip}.

Anyhow, in the integral form as in Equation~(\ref{eq:int_drVP}) it is clear that it is the radial derivative of the vacuum polarization, on both sides of the horizon that determines locally the contribution to the~depletion.

%---------------------------------------------------------------------
\section{Discussion and Conclusions} 
%---------------------------------------------------------------------
We have studied the canonical analogue black hole as a particular realization of analogue gravity with Bose--Einstein condensates.

The linearized quantum fluctuation over the background---the mean-field described by the Gross--Pitaevskii equation---propagates as a massless scalar quantum field in the curved spacetime of a black hole.
In the limit of negligible quantum pressure, the Bogoliubov--de Gennes equations can be reorganized to show how the quantum fluctuation of the phase is governed by an acoustic metric, defined by the condensate wavefunction, which presents a horizon. The calculation of the Hawking radiation produced in this acoustic geometry follows closely that of a massless scalar field near a gravitational black hole, and we have set the conditions for the modes of the field to replicate the Unruh vacuum, the state expected in a black hole spacetime formed through gravitational collapse.

As we are interested in the back-reaction that the quasi-particles exert on the condensate, we have modified the Gross--Pitaevskii equation to include the anomalous terms, and thus observed how the acoustic metric is modified by the quasi-particle dynamics. The study of the quasi-particle back-reaction 
%pushes analogue gravity towards an understanding of semiclassical gravity, where the back-reaction of the quantum fields is expected to be included in Einstein's equations. 
%
%Such a study 
{in analogue models can hopefully provide a much needed insight for understanding how, in an emergent gravity scenario, the evaporation process could be described beyond semiclassical gravity, at the fundamental level.~(Note however that transplanckian fluctuations can be accounted for in a semiclassical treatment of the back-reaction, as for example it is discussed in~\cite{PhysRevD.54.7444}.)

We have provided expressions for the back-reaction in analogue gravity with Bose--Einstein condensates, in an approximation based on the suppression of the terms depending on the healing length scale---subleading with respect to scale set by the Hawking temperature---that leads to an equation that focuses on the vacuum polarization of the analogue scalar field.

From this general result, we have specialized to the canonical analogue black hole and have obtained an expression for the dynamics of the horizon and for how it is affected by the state of the quantum field. In particular, we have argued that in a regime in which the Hawking radiation is the leading contribution in the dynamics of the acoustic geometry (that would otherwise be stationary) it leads to a shrinking of the horizon radius.

This process of evaporation induces the depletion of atoms from the condensate to the excited part, and the rate of depletion is driven by the radial derivative of the vacuum polarization. The~exchange of atoms and therefore the exchange of information between the two sectors of the system happens not only at the horizon, but also in the black hole region enclosed by it. In particular, we have obtained an expression which ties together the rate of depletion to the number of atoms in the region within the horizon, the temperature of the black hole and the value of the vacuum polarization near the singularity. These are all results in agreement with the conclusions reported in~\cite{Liberati:2019fse} with regards to the nature of a possible resolution of the information loss problem associated with black hole evaporation. In~particular, there we conjectured that the number of atoms in the region of the condensate enclosed by the horizon is the relevant quantity for the suppression of the correlators between condensate atoms and Hawking quanta; correlators which are needed for preserving unitarity during evaporation. This~is corroborated by our finding that the back-reaction acts through the depletion of the atoms in the black hole region.
 
What is peculiar of the presented calculation is the role played by the vacuum polarization in the inner region. In the gravitational case of the Schwarzschild black hole there is neither access to the information in the inner region nor knowledge of an underlying quantum gravity structure from which the classical geometry would emerge. Instead, in the canonical analogue black whole there are both these features. Not only in an experimental realization one has access to the whole space involved in the dynamics, but it is now clear that the role played by the vacuum polarization of the analogue field, the Bogoliubov quasi-particles, is fundamental in understanding the exchange of information between the condensate and the excited part of the system. 

We can then conclude that a deeper understanding of the dynamics of acoustic horizons in BEC-based analogue gravity, and in particular of the associated phenomenology of depletion, could~give---despite the different form of the gravitodynamic equations---precious insights towards a deeper understanding of the role of back-reaction in semiclassical gravity in the entanglement between quantum matter and quantum spacetime degrees of freedom. An entanglement which might be key for understanding the compatibility Hawking radiation with unitarity and hence for understanding black holes as thermodynamic objects.\\

{\em Acknowledgements:} {Very useful discussions and correspondence with M. Visser are gratefully acknowledged. SL acknowledges funding from the Ministry of Education and Scientific Research (MIUR) under the grant PRIN MIUR 2017-MB8AEZ.}

%
%
%%---------------------------------------------------------------------
%\section*{Acknowledgments} Acknowledgements here
%---------------------------------------------------------------------

%\appendix
%\addappheadtotoc
%\appendixpage
%---------------------------------------------------------------------
\section*{References}
%---------------------------------------------------------------------
%\begin{thebibliography}{69}  
%======================================================

\bibliographystyle{unsrt}
\bibliography{BEC_refs}{}

\end{document}